\documentclass[prd,showpacs,showkeys,floatfix,twocolumn,amsmath,amssymb,floatfix]{revtex4}
\usepackage{graphicx,color,dcolumn,booktabs,bm}
\usepackage{longtable,lscape}
\usepackage{amssymb}
\usepackage{indentfirst}
\usepackage{epsfig}
\usepackage{feynmf}   
\usepackage{epstopdf}   
\usepackage{slashed}  
\usepackage{cases}
\usepackage{color}
\usepackage{multirow}
\usepackage{graphicx,color,dcolumn,booktabs,bm}
\usepackage[colorlinks, citecolor=blue,anchorcolor=red,menucolor=red, linkcolor=red,filecolor=red,runcolor=red,urlcolor=blue,frenchlinks=red]{hyperref}

\begin{document}

\title{$F$-wave heavy-light meson spectroscopy in QCD sum rules and heavy quark effective theory}

\author{Dan Zhou}
\affiliation{
School of Physics and Nuclear Energy Engineering and International Research Center for Nuclei and Particles in the Cosmos, Beihang University, Beijing 100191, China
}
\author{Hua-Xing Chen}
\email{hxchen@buaa.edu.cn}
\affiliation{
School of Physics and Nuclear Energy Engineering and International Research Center for Nuclei and Particles in the Cosmos, Beihang University, Beijing 100191, China
}
\author{Li-Sheng Geng}
\affiliation{
School of Physics and Nuclear Energy Engineering and International Research Center for Nuclei and Particles in the Cosmos, Beihang University, Beijing 100191, China
}
\author{Xiang Liu}
\email{xiangliu@lzu.edu.cn}
\affiliation{
School of Physical Science and Technology, Lanzhou University, Lanzhou 730000, China \\
Research Center for Hadron and CSR Physics, Lanzhou University and Institute of Modern Physics of CAS, Lanzhou 730000, China
}
\author{Shi-Lin Zhu}
\email{zhusl@pku.edu.cn}
\affiliation{
School of Physics and State Key Laboratory of Nuclear Physics and Technology, Peking University, Beijing 100871, China \\
Collaborative Innovation Center of Quantum Matter, Beijing 100871, China \\
Center of High Energy Physics, Peking University, Beijing 100871, China
}

\begin{abstract}
We study the $F$-wave $\bar c s$ heavy meson doublets $(2^+,3^+)$
and $(3^+,4^+)$. They have large orbital excitations $L=3$, and may be good
challenges (tests) for theoretical studies. To study them we use the method of QCD sum rule in the
framework of heavy quark effective theory. Their masses are predicted
to be $m_{(2^+,3^+)}$ = ($3.45 \pm 0.25$, $3.50 \pm 0.26$) GeV and
$m_{(3^+,4^+)}$ = ($3.20 \pm 0.22$, $3.26 \pm 0.23$) GeV, with mass splittings $\Delta m_{(2^+,3^+)} = m_{3^+} - m_{2^+} =  0.046 \pm 0.030$ GeV and $\Delta m_{(3^+,4^+)}
=  0.053 \pm 0.044$ GeV, respectively. We note that this is a pioneering work and these results are provisional.
\end{abstract}

\maketitle

\section{Introduction}
\label{sec:intro}

Since 2003, big progress on the observations of heavy-light mesons has been made. When checking 2014 edition of Particle Data Group (PDG) \cite{Agashe:2014kda}, we notice that charmed meson, charmed-strange meson, bottom meson, and bottom-strange meson families have become more and more abundant, which is due to these observed candidates of higher radial and orbital excitations of heavy-light mesons.
In the following years, theorists and experimentalists will pay more attentions to the study of heavy-light mesons with higher radial and orbital quantum numbers, especially with the running of LHCb and forthcoming BelleII.

With the charmed-strange meson family as example,  we introduce the research status of higher excitations of heavy-light meson.
There are two $1S$ states ($D_s(1968)$ and $D_s^*(2112)$) and four $1P$ states ($D_{s0}^*(2317)$, $D_{s1}(2460)$, $D_{s1}(2536)$, and $D_{s2}^*(2573)$) established in PDG \cite{Agashe:2014kda}. The observed $D_{s1}^*(2700)$ \cite{Aubert:2006mh}, $D_{s1}^*(2860)$ \cite{Aaij:2014xza,Aaij:2014baa}, and $D_{s3}^*(2860)$ \cite{Aaij:2014xza,Aaij:2014baa} stimulated theorist's interest in studying the properties of $2S$ and $1D$ states \cite{Zhang:2006yj,Song:2015nia,Segovia:2015dia}, while the observation of $D_{sJ}(3040)$ \cite{Aubert:2009ah} made
us to focus on the $2P$ states \cite{Sun:2009tg}. The research status of charmed-strange mesons can be found by a mini review \cite{Liu:2010zb} and two recent systematical theoretical work \cite{Song:2015nia,Segovia:2015dia}.
The theoretical and experimental situation of charmed meson is similar to that of charmed-strange meson, which can be found in Ref. \cite{Song:2015fha}.

Considering the above research status of heavy-light meson, it is suitable time to carry out the study of F-wave heavy-light mesons, since these 1F states will be reported in future experiment. The calculation of mass spectrum of $F$-wave heavy-light mesons can provides valuable information to experimental search for them.
Before the present work, there were several quark model calculation of mass spectrum of $F$-wave heavy-light mesons.
For example, Ebert {\it et al.} adopted the relativistic quark model to get the heavy-light meson spectroscopy \cite{Ebert:2009ua}, which includes the $1F$ states. In Ref. \cite{Di Pierro:2001uu}, a relative quark model including the leading order correction in $1/m_{c,b}$
was applied to study heavy-light meson masses and light hadronic transition rates, where this study also contains $1F$ states.
Recently, in Refs. \cite{Song:2015nia,Song:2015fha}, the masses of $1F$ states in charmed meson and charmed-strange meson families were obtained through the modified Godfrey-Isgur (GI) model, where the screening effect is considered in the introduced potential. For bottom and bottom-strange mesons,
the masses of the $1F$ states were estimated by the GI model in Ref. \cite{Sun:2014wea}.

Although there were quark model calculations of $1F$ states of heavy-light mesons.
we notice that a QCD sum rule (QSR) study of mass spectrum of $F$-wave heavy-light mesons is still absent at present, which inspires our interest in performing the calculation of QSR of mass spectrum of $F$-wave heavy-light mesons. In Refs.~\cite{Shifman:1993wf,Shifman:1982zt} M.~A.~Shifman wrote about QSR that:
$\\$
``{\it One failure is quite obvious: the large-spin
hadrons. Indeed, the latter have parametrically large sizes and a
`sausage-like shape' (growing with spin) and, therefore, it is quite
clear that the basic idea of the method -- extrapolation from short to
intermediate distances -- is not applicable. Practically, we have to
stop at S=2.}''
$\\$
However, it is still worth a try to applying QSR to study $F$-wave heavy mesons, because a) we have used the same method to
well study $D$-wave heavy mesons~\cite{Zhou:2014ytp} and
$P$-wave heavy baryons~\cite{Chen:2015kpa}; and b) the LHCb experiments have just observed $D$-wave heavy mesons~\cite{Aaij:2014xza,Aaij:2014baa},
and $F$-wave heavy mesons are expected in the following experiments. Hence, the present pioneering study not only provides important hint to experimental
exploration of $F$-wave heavy-light mesons, but also test the applicability of QSR when applying QSR to study so higher radial excitations.
This can be useful for quantifying potential overextensions of QSR in order to inspire ideas for its improvement, especially
with future experimental data on $F$-wave heavy mesons.

The $F$-wave $\bar Q s$ ($Q=c,b$) heavy mesons have large orbital excitations $L=3$, and may be good
challenges (tests) for theoretical studies. Based on the heavy quark effective theory
(HQET)~\cite{Grinstein:1990mj,Eichten:1989zv,Falk:1990yz}, we can classify them into two doublets, $(2^+,3^+)$ and $(3^+,4^+)$, the light components
of which have $j_l^{P_l} = 5/2^-$ and $j_l^{P_l} = 7/2^-$, respectively.
In this paper we shall use the method of QCD sum rule~\cite{Shifman:1978bx,Reinders:1984sr} to study them,
which has been successfully applied to study the ground state ($S$-wave) heavy meson doublet $(0^-,1^-)$~\cite{Bagan:1991sg,Neubert:1991sp,Neubert:1993mb,Broadhurst:1991fc,Ball:1993xv,Huang:1994zj,Colangelo:1991ug,Colangelo:1992kc},
the $P$-wave heavy meson doublets $(0^+,1^+)$ and $(1^+,2^+)$~\cite{Dai:1996yw,Dai:1993kt,Dai:1996qx,Dai:2003yg,Colangelo:1998ga},
and the $D$-wave heavy meson doublets $(1^-,2^-)$ and $(2^-,3^-)$~\cite{Zhou:2014ytp}. In this paper
we shall follow the procedures used in these references, and study
the $F$-wave $\bar c s$ heavy meson doublets $(2^+,3^+)$ and
$(3^+,4^+)$. In the calculations we shall take into account the ${\mathcal O}(1/m_Q)$ corrections, where $m_Q$ is the
heavy quark mass. We note that the convergence of this $1/m_Q$ expansion can be problematic because $F$-wave heavy mesons (probably)
have masses significantly larger than the heavy quark mass. However, we still hope that the leading
terms and the ${\mathcal O}(1/m_Q)$ corrections could capture sufficiently much of the most important qualitative
physics. We shall also carefully check this convergence in Sec.~\ref{sec:summary}.

This paper is organized as follows. After this Introduction, we
construct the $F$-wave $\bar c s$ interpolating currents for the
heavy meson doublets $(2^+,3^+)$ and $(3^+,4^+)$ in Sec.~\ref{sec:leading}.
These currents are then used to
perform QCD sum rule analyses in the framework of HQET both at the leading order and
at the ${\mathcal O}(1/m_Q)$ order. The calculations are
done in Sec.~\ref{sec:leading} and Sec.~\ref{sec:nexttoleading}, and
the results are summarized in Sec.~\ref{sec:summary}.

\section{The Sum Rules at the Leading Order (in the $m_Q \rightarrow \infty$ limit)}
\label{sec:leading}

The heavy meson interpolating currents have been systematically constructed in Refs.~\cite{Dai:1993kt,Dai:1996yw,Dai:1996qx}. Here we follow Ref.~\cite{Zhou:2014ytp} and briefly show how we construct the $F$-wave interpolating currents. We denote them as $J^{\alpha_1\cdots\alpha_{j}}_{j,P,j_l}$, where $j$ and $P$ are the total angular momentum and parity of the heavy meson, and $j_l$ is the total angular momentum of the light components (containing three orbital excitations). We have the following relation
\begin{eqnarray}
\vec{j} = \vec j_l \otimes \vec s_Q \, ,
\end{eqnarray}
where $s_Q = 1/2$ is the spin of the heavy quark.

To construct the $F$-wave interpolating currents, we just need to add three derivatives to the pseudoscalar current $\bar h_v \gamma_5 q$ of $J^P=0^-$ and the vector current $\bar h_v \gamma_\mu q$ of $J^P=1^-$. By doing this, the three orbital excitations can be explicitly written up. We act them on the light (strange) quark, and the obtained field has either $j_l^{P_l} = 5/2^-$:
\begin{eqnarray}
\label{eq:52}
\mathcal{D}_t^{\alpha_1} \mathcal{D}_t^{\alpha_2} \mathcal{D}_t^{\beta} \times \gamma_\beta \gamma_5 q \, ,
\end{eqnarray}
or $j_l^{P_l} = 7/2^-$:
\begin{eqnarray}
\label{eq:72}
\mathcal{D}_t^{\alpha_1} \mathcal{D}_t^{\alpha_2} \mathcal{D}_t^{\alpha_3} \times q \, ,
\end{eqnarray}
where $D^\mu_t = D^\mu - (D \cdot v) v^\mu$ with $D^\mu = \partial^\mu - i g A^\mu$. Some other notations are: $\gamma_t^\mu = \gamma^\mu - v\!\!\!\slash v^\mu$,  $h_v$ denotes the heavy quark field in HQET, $v$ is the velocity of the heavy quark, and $g_t^{\alpha_1\alpha_2}=g^{\alpha_1\alpha_2} - v^{\alpha_1} v^{\alpha_2}$ denotes the transverse metric tensor.

We use Eq.~(\ref{eq:52}) of $j_l^{P_l} = 5/2^-$ to construct the interpolating currents coupling to the $F$-wave $(2^+,3^+)$ spin doublet, based on $\bar h_v \gamma_5 q$ and $\bar h_v \gamma_\mu q$:
\begin{eqnarray}
J^{\dag \alpha_1 \alpha_2}_{x,+,5/2} &=& \bar h_v \gamma_5 \times \mathcal{D}_t^{\alpha_1} \mathcal{D}_t^{\alpha_2} \mathcal{D}_t^{\beta} \times \gamma_\beta \gamma_5 q \, ,
\\ J^{\dag \alpha_1 \alpha_2 \alpha_3}_{y,+,5/2} &=& \bar h_v \gamma_t^{\alpha_3} \times \mathcal{D}_t^{\alpha_1} \mathcal{D}_t^{\alpha_2} \mathcal{D}_t^{\beta} \times \gamma_\beta \gamma_5 q \, .
\end{eqnarray}
Here $x$ and $y$ mean that these two currents are not pure $2^+$ nor $3^+$, while we can project out the two pure ones:
\begin{eqnarray}
J^{\dag \alpha_1 \alpha_2}_{2,+,5/2} &=& \sqrt{\frac{5}{6}} \bar h_v (-i)^3 \mathcal{S}_2 \big[ \mathcal{D}_t^{\alpha_2} (\mathcal{D}_t^{\alpha_1} - \frac{2}{5} \gamma_t^{\alpha_1} \slashed D_t) \slashed D_t \big] q \, , \nonumber\\\label{eq:current1}
\\ J^{\dag \alpha_1 \alpha_2 \alpha_3}_{3,+,5/2} &=& \sqrt{\frac{1}{2}} \bar h_v \gamma^5 {(-i)^3} \mathcal{S}_3 \big[ \gamma_t^{\alpha_1} \mathcal{D}_t^{\alpha_2} \mathcal{D}_t^{\alpha_3} \slashed D_t \big] q \, ,
\label{eq:current2}
\end{eqnarray}
where $\mathcal{S}_j$ denotes symmetrization and subtracting the trace terms in the sets $(\alpha_1 \cdots \alpha_j)$.
We note that the expressions of these currents have been modified to be consistent with Refs.~\cite{Dai:1993kt,Dai:1996yw,Dai:1996qx}.

Similarly, we use Eq.~(\ref{eq:72}) of $j_l^{P_l} = 7/2^-$ to construct the interpolating currents coupling to the $F$-wave $(3^+,4^+)$ spin doublet:
\begin{eqnarray}
J^{\dag \alpha_1 \alpha_2 \alpha_3}_{3,+,7/2} &=& \sqrt{\frac{7}{8}} \bar h_v \gamma_5 (-i)^3 \mathcal{S}_3 \big[ \mathcal{D}_t^{\alpha_2} \mathcal{D}_t^{\alpha_3} (\mathcal{D}_t^{\alpha_1}\nonumber\\&& - \frac{3}{7} \gamma_t^{\alpha_1} \slashed D_t) \big] q \, , \label{eq:current3}
\\ J^{\dag \alpha_1 \alpha_2 \alpha_3 \alpha_4}_{4,+,7/2} &=& \sqrt{\frac{1}{2}} \bar h_v {(-i)^3} \mathcal{S}_4 \big[ \gamma_t^{\alpha_1} \mathcal{D}_t^{\alpha_2} \mathcal{D}_t^{\alpha_3} \mathcal{D}_t^{\alpha_4} \big] q \, .
\label{eq:current4}
\end{eqnarray}
These interpolating currents are then used to perform QCD sum rule analyses.
As discussed in Refs.~\cite{Dai:1993kt,Dai:1996yw,Dai:1996qx}, we do not need to
investigate all of them, but just choose $J^{\dag \alpha_1 \alpha_2 \alpha_3}_{3,+,5/2}$ and $J^{\dag \alpha_1 \alpha_2 \alpha_3 \alpha_4}_{4,+,7/2}$, because the calculation using these two currents are
a bit simpler (to be technically precise, we use non-symmetrized currents to calculate the operator product expansion (OPE)
and then do the ``symmetrization and subtracting the trace terms''). Moreover, we shall fix $q$ to be the strange quark in the following,
because we are mainly studying $\bar c s$ heavy mesons in this paper.

We follow the procedures used in Ref.~\cite{Zhou:2014ytp}, and assume
$|j,P,j_l\rangle$ to be the heavy meson state with the quantum
numbers $j, P$ and $j_l$ in the $m_Q \rightarrow \infty$ limit.
The relevant interpolating field couples to it through
\begin{eqnarray}
\label{eq:defg}
\langle 0| J^{\alpha_1\cdots\alpha_j}_{j,P,j_l}
|j^\prime,P^\prime,j_l^\prime \rangle = f_{P,j_l} \delta_{j
j^\prime} \delta_{P P^\prime} \delta_{j_l j_l^\prime}
\eta^{\alpha_1\cdots\alpha_j}_t \, ,
\end{eqnarray}
where $f_{P, j_l}$ denotes the decay constant, and $\eta^{\alpha_1\cdots\alpha_j}_t$ denotes the transverse, traceless, and symmetric polarization tensor, satisfying:
\begin{eqnarray}
\eta^{\alpha_1 \cdots \alpha_j}_t \eta^{*\beta_1 \cdots \beta_j}_t &=& \mathcal{S}^\prime_j [\tilde g^{\alpha_1 \beta_1}_t \cdots \tilde g^{\alpha_j \beta_j}_t] \, .
\end{eqnarray}
In this expression $\tilde g_t^{\mu \nu} = g_t^{\mu \nu} - q_t^\mu q_t^\nu / m^2$, and $\mathcal{S}^\prime_j$ denotes symmetrization and subtracting the trace
terms in the sets $(\alpha_1 \cdots \alpha_j)$ and $(\beta_1 \cdots \beta_j)$. Based on Eq.~(\ref{eq:defg}), we can construct the two-point correlation function
\begin{eqnarray}
&&\Pi^{\alpha_1\cdots\alpha_j,\beta_1\cdots\beta_j}_{j,P,j_l} (\omega)
\nonumber\\&&= i \int d^4 x e^{i k x}\left \langle 0\left | T\left[J^{\alpha_1\cdots\alpha_j}_{j,P,j_l}(x) J^{\dagger\beta_1\cdots\beta_j}_{j,P,j_l}(0)\right] \right| 0 \right\rangle
\nonumber
\\  &&= (-1)^j \mathcal{S}^\prime_j \left[ \tilde g_t^{\alpha_1 \beta_1} \cdots \tilde g_t^{\alpha_j \beta_j} \right] \Pi_{j,P,j_l} (\omega) \, ,
\label{eq:pi}
\end{eqnarray}
and calculate it at the hadron level:
\begin{eqnarray}
\Pi_{j,P,j_l}(\omega) &=& {f_{P,j_l}^2 \over 2 \bar \Lambda_{P,j_l} - \omega} + \mbox{higher states} \, ,
\label{eq:pole}
\end{eqnarray}
where $\omega = 2 v \cdot k$ denotes twice the external off-shell energy. $\bar \Lambda_{P,j_l} = \bar \Lambda_{j_l-1/2,P,j_l} = \bar \Lambda_{j_l+1/2,P,j_l}$ is defined to be
\begin{eqnarray}
\label{eq:lambda}
\bar \Lambda_{P,j_l} &\equiv& \lim_{m_Q \rightarrow \infty} (m_{j,P,j_l} - m_Q) \, ,
\end{eqnarray}
where $m_{j,P,j_l}$ is the mass of the lowest-lying state which $J^{\alpha_1\cdots\alpha_j}_{j,P,j_l}(x)$ couples to.

We can also calculate Eq.~(\ref{eq:pi}) at the quark
and gluon level using the method of QCD sum rule in the
framework of the heavy quark effective theory, i.e., we insert
Eq.~(\ref{eq:current2}) and (\ref{eq:current4}) into
Eq.~(\ref{eq:pi}), perform the Borel transformation, and then
obtain (see Refs.~\cite{Dai:1993kt,Dai:1996yw,Dai:1996qx,Dai:2003yg,Zhou:2014ytp} for details):
\begin{eqnarray}
&&\Pi_{3,+,5/2}(\omega_c, T) = f_{+,5/2}^2 e^{-2 \bar \Lambda_{+,5/2} / T}
\nonumber\\  \quad&&=
\int_{2 m_s}^{\omega_c} \bigg[ {3 \over 17920 \pi^2} \omega^8 + {3 m_s \over 8960 \pi^2}  \omega^7 - {3 m_s^2 \over 1280 \pi^2} \omega^6 \nonumber\\&&\quad- {{\langle g_s^2 GG \rangle} \over 144 \pi^2} \omega^4 \bigg] e^{-\omega/T} d\omega \, ,
\label{eq:ope1}
\\ &&\Pi_{4,+,7/2}(\omega_c, T) = f_{+,7/2}^2 e^{-2 \bar \Lambda_{+,7/2} / T}
\nonumber\\  \quad&&=
\int_{2 m_s}^{\omega_c} \bigg[ {3 \over 17920 \pi^2} \omega^8 + {3 m_s \over 8960 \pi^2}  \omega^7 - {3 m_s^2 \over 1280 \pi^2} \omega^6 \nonumber\\&&\quad- {{19 \langle g_s^2 GG \rangle} \over 3072 \pi^2} \omega^4 \bigg] e^{-\omega/T} d\omega \, .
\label{eq:ope2}
\end{eqnarray}
These two sum rules for $(2^+,3^+)$ and $(3^+,4^+)$ are similar.
Similarly to Ref.~\cite{Zhou:2014ytp},
the quark condensate $\langle \bar q q \rangle$ and
the mixed condensate $\langle g_s \bar q \sigma G q \rangle$ both
vanish, making the convergence of Eqs.~(\ref{eq:ope1}) and (\ref{eq:ope2})
quite good. This can be easily verified because we need to apply as many
as six covariant derivatives to the light quark propagator
%
\begin{eqnarray}
\label{eq:propagator}
&& i\mbox{S}^{ab}_q(y,x) \equiv \langle0|\mbox{T}[q^a(y)\bar{q}^b(x)]|0\rangle
\\ \nonumber &=& \frac{i\delta^{ab}(\hat{y} - \hat{x})}{2\pi^2(y-x)^4}
-\frac{\delta^{ab}}{12}\langle\bar{q}q\rangle
-\frac{\delta^{ab}(y-x)^2}{192}\langle g_c\bar{q}\sigma Gq\rangle
\\ \nonumber &-& \frac{\delta^{ab}m_q}{4\pi^2(y-x)^2}
+\frac{i\delta^{ab}m_q }{48}\langle\bar{q}q\rangle(\hat{y} - \hat{x})
+\frac{i\delta^{ab}(\hat{y} - \hat{x})m_q^2}{8\pi^2(y-x)^2}
\\ \nonumber &-& \frac{i}{32\pi^2}\frac{\lambda^n_{ab}}{2}\mbox{g}_c\mbox{G}^n_{\mu\nu}\frac{1}{(y-x)^2}(\sigma^{\mu\nu}(\hat{y} - \hat{x}) + (\hat{y} - \hat{x})\sigma^{\mu\nu})
\\ \nonumber &+& \frac{1}{4\pi^2}\frac{\lambda^n_{ab}}{2}\mbox{g}_c\mbox{G}^n_{\mu\nu}\frac{1}{(y-x)^4}(\hat{y} - \hat{x}) y_{\mu} x_{\nu} \, .
\end{eqnarray}
%
Differently, we need to carefully deal with the
gluon terms contained in these covariant derivatives in order to evaluate the gluon condensate,
which gives significant contribution.
The gluon condensate and
the strange quark mass take the following values~\cite{Dai:1993kt,Dai:1996yw,Dai:1996qx,Dai:2003yg,Ioffe:2005ym}:
%
\begin{eqnarray}
\langle {\alpha_s\over\pi} GG\rangle &=& 0.005 \pm 0.004 \mbox{ GeV}^4\, ,
\\ \label{condensates}  m_s &=& 0.15 \mbox{ GeV} \, .
\end{eqnarray}

We note that the radiative corrections are not taken into account in our calculations,
which can be important but not easy to evaluate, because the six covariant derivatives also contribute to them
(see discussions related to $f_B$ in Ref.~\cite{Narison:1987qc} and related references).
However, we expect that they would lead
an uncertainty significantly smaller than the gluon condensate and the charm quark mass.
Hence, we shall discuss the change of the latter two parameters in Sec.~\ref{sec:summary},
but do not discuss the radiative corrections any more.

To obtain $\bar \Lambda_{P,j_l}$, we just need to differentiate Log[Eq.~(\ref{eq:ope1})] and Log[Eq.~(\ref{eq:ope2})]
with respect to $[-2/T]$:
\begin{eqnarray}
\bar \Lambda_{P,j_l}(\omega_c, T) &=& \frac{\partial[{\rm Log}\Pi_{j,P,j_l}(\omega_c,T)]}{\partial[-2/T]} \, . \label{eq:mass}
\end{eqnarray}
Then we can use it to further evaluate $f_{P,j_l}$:
\begin{eqnarray}
f_{P,j_l}(\omega_c, T) &=& \sqrt{e^{2 \bar \Lambda_{P,j_l}(\omega_c, T) / T} \times \Pi_{j,P,j_l}(\omega_c, T)} \, .
\label{eq:coupling}
\end{eqnarray}
There are two free parameters in these equations: the
Borel mass $T$ and the threshold value $\omega_c$. We need to fix
these two parameters to evaluate $\bar \Lambda_{P,j_l}(\omega_c, T)$
and $f_{P,j_l}(\omega_c, T)$.

\begin{figure}[hbtp]
\begin{center}
\begin{tabular}{c}
\scalebox{0.6}{\includegraphics{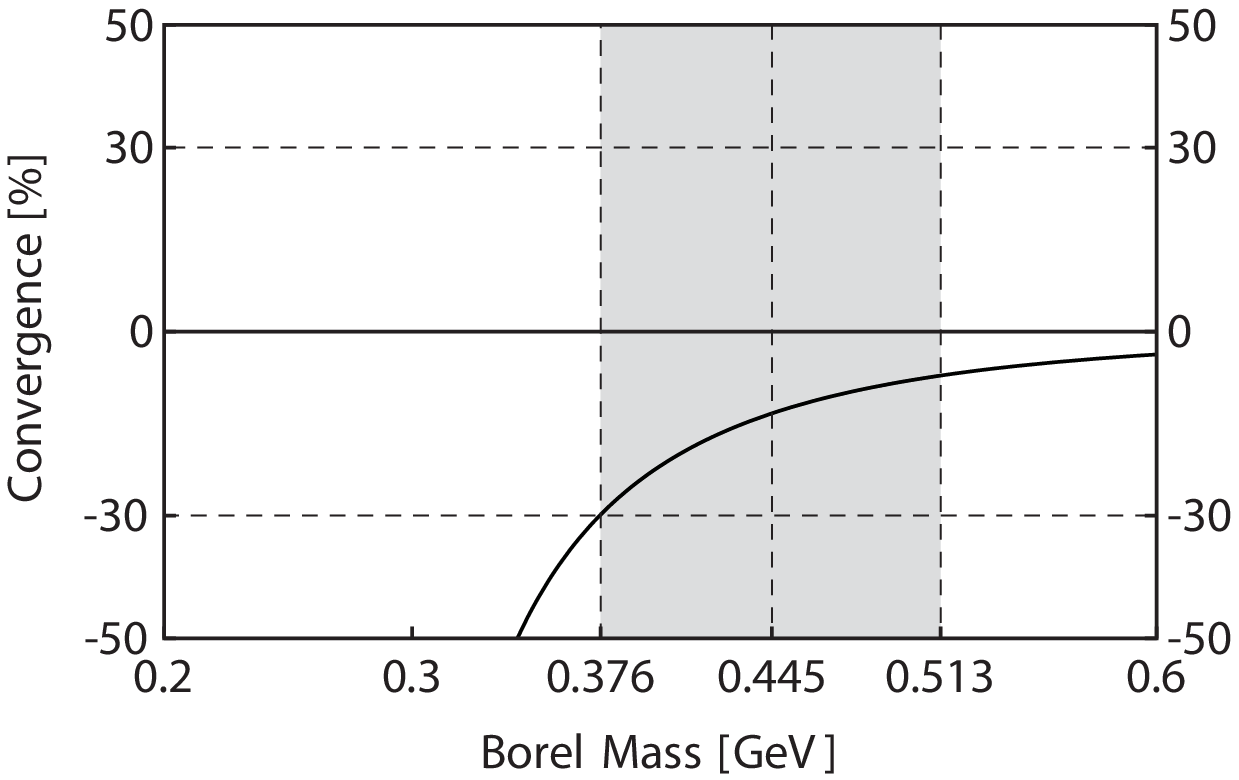}}\\
\scalebox{0.573}{\includegraphics{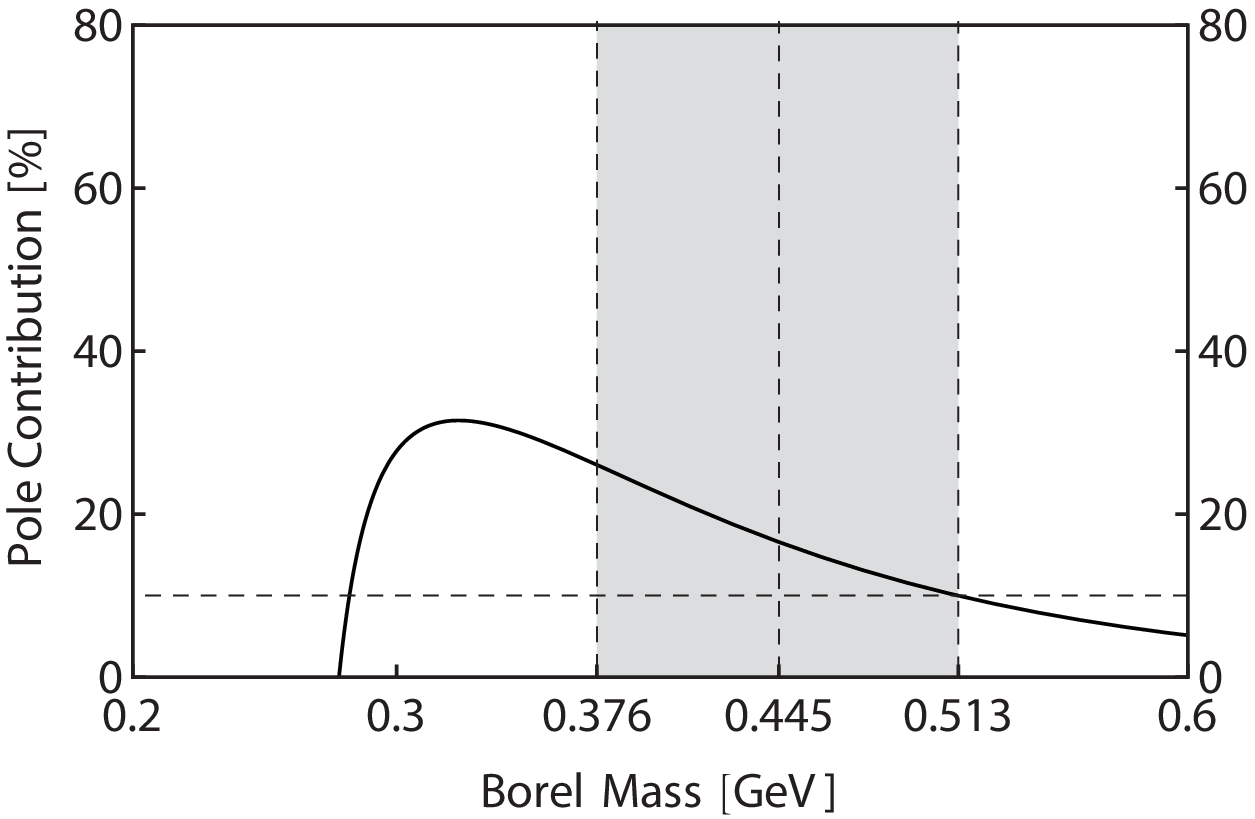}}
\end{tabular}
\caption{The variations of CVG and PC with respect to the Borel mass $T$.
The sum rule (\ref{eq:ope1}) for the current $J^{\dag \alpha_1 \alpha_2 \alpha_3}_{3,+,5/2}$ is used in both figures.}
\label{fig:criteria}
\end{center}
\end{figure}

We use two criteria to fix the Borel mass $T$. One criterion is to require that the high-order power corrections be less than 30\%
to determine its lower limit $T_{min}$:
%
\begin{equation}
\label{eq_convergence}
\mbox{Convergence (CVG)} \equiv \left|\frac{ \Pi^{\rm high-order}_{j,P,j_l}(\infty, T) }{ \Pi_{j,P,j_l}(\infty, T) }\right| \leq 30\% \, ,
\end{equation}
%
where $\Pi^{\rm high-order}_{j,P,j_l}(\omega_c, T)$ denotes the high-order power corrections, for example,
%
\begin{eqnarray}
\Pi^{\rm high-order}_{3,+,5/2}(\omega_c, T) &=& \int_{2 m_s}^{\omega_c} \left[ - {{\langle g_s^2 GG \rangle} \over 144 \pi^2} \omega^4  \right] e^{-\omega/T} d\omega \, .
\nonumber \\
\end{eqnarray}
%
The other criterion is to require that the pole contribution (PC) be larger than 10\% to determine its upper limit $T_{max}$:
%
\begin{equation}
\label{eq_pole} \mbox{PC} \equiv \frac{ \Pi_{j,P,j_l}(\omega_c, T) }{ \Pi_{j,P,j_l}( \infty , T) } \geq 10\% \, .
\end{equation}
%
Altogether we obtain a Borel window $T_{min}<T<T_{max}$ for a fixed threshold value $\omega_c$. This $\omega_c$ is the
other free parameter, which will be fixed in Sec.~\ref{sec:summary}.
Here we proceed using the sum rule (\ref{eq:ope1}) and taking $\omega_c = 3.0$ GeV
as an example. Using this value of $\omega_c$, we obtain
a Borel window $0.376$ GeV $< T < 0.513$ GeV for the sum rule (\ref{eq:ope1}):
the lower limit is determined by using the first criterion of CVG, as shown in
the top panel of Fig.~\ref{fig:criteria}, and the upper limit is determined by using the second
criterion of PC, as shown in the bottom panel of Fig.~\ref{fig:criteria}.

\begin{figure}[hbtp]
\begin{center}
\begin{tabular}{c}
\scalebox{0.6}{\includegraphics{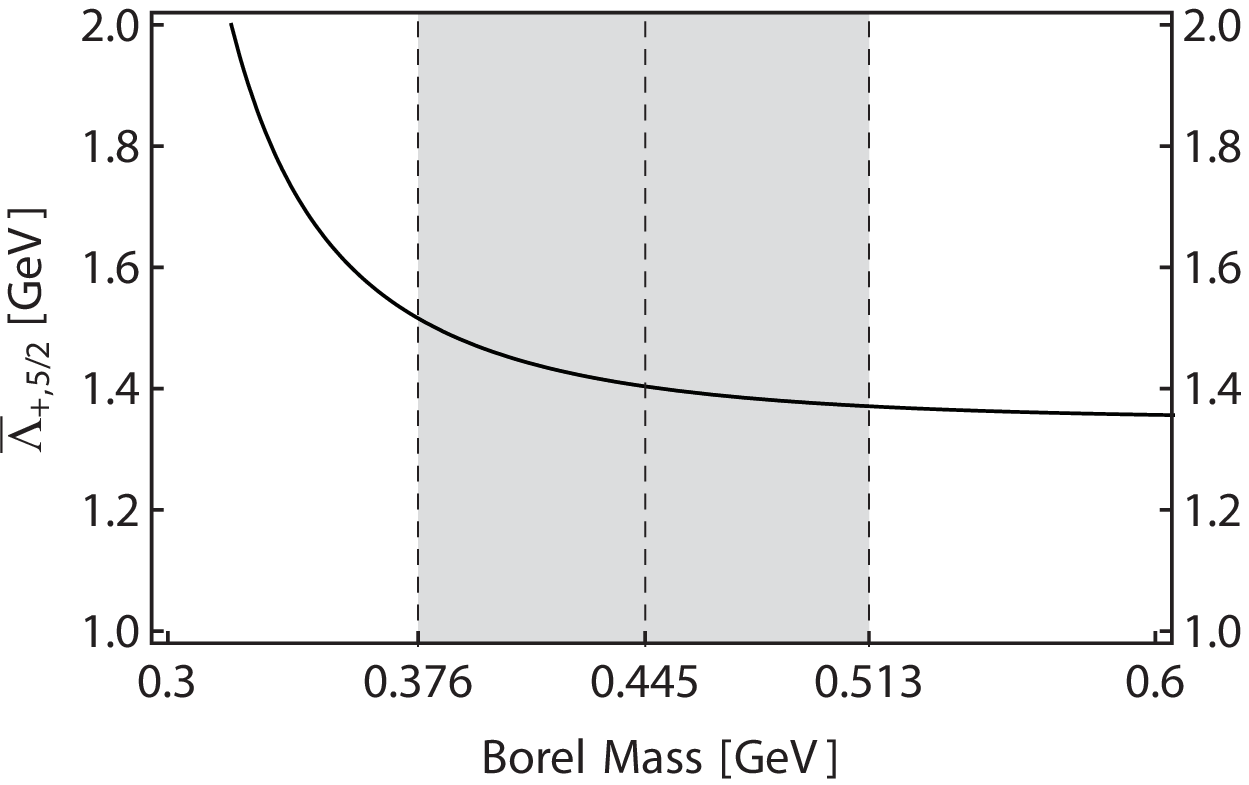}}\\
\scalebox{0.6}{\includegraphics{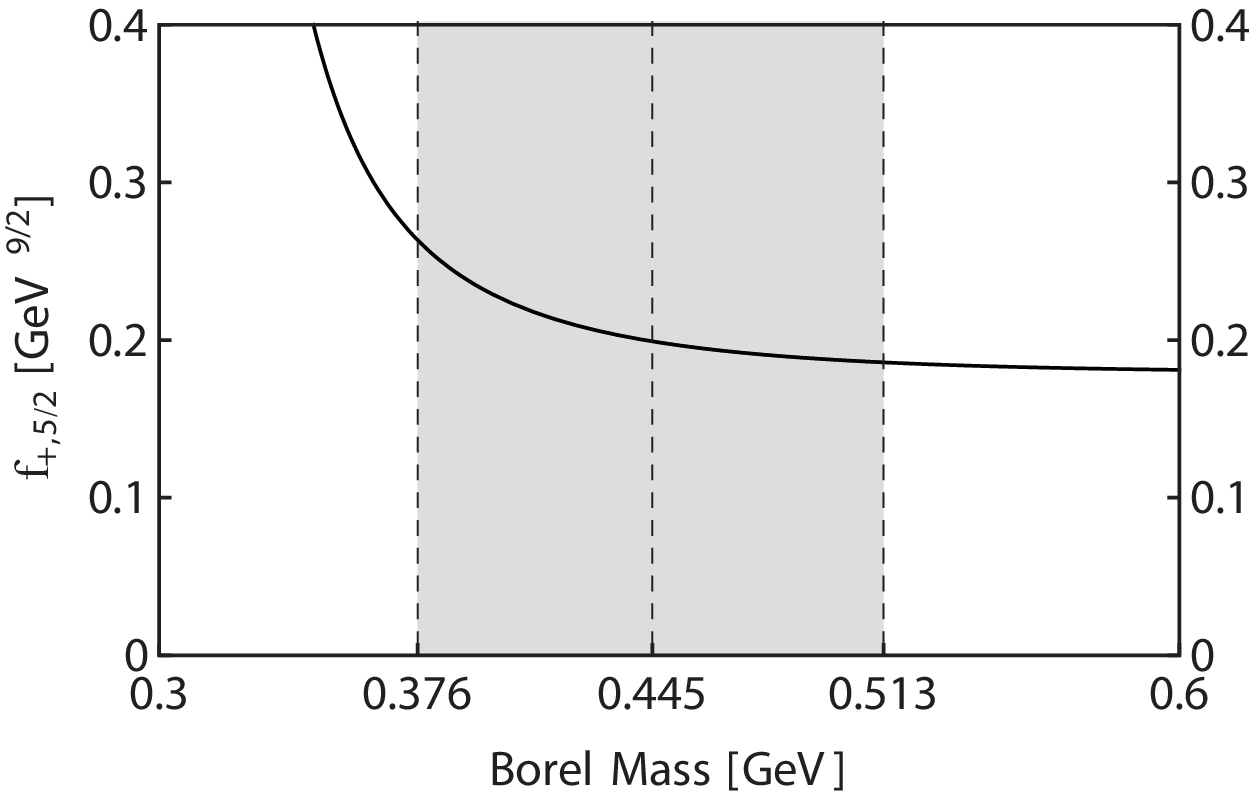}}
\end{tabular}
\caption{The
variations of $\bar \Lambda_{+,5/2}$ (top) and $f_{+,5/2}$ (bottom) with respect to the
Borel mass $T$.
In both figures we take $\omega_c$ = 3.0 GeV and the Borel window is $0.376$ GeV $< T < 0.513$ GeV.}
\label{fig:lambda1}
\end{center}
\end{figure}

Finally, we show the variations of $\bar \Lambda_{+,5/2}$ and $f_{+,5/2}$ with respect to the
Borel mass $T$ in Fig.~\ref{fig:lambda1}. We show them
in a broader region $0.3$ GeV $< T < 0.6$ GeV, while these curves are more stable in the Borel
window $0.376$ GeV $< T < 0.513$ GeV. We obtain the following numerical results:
\begin{eqnarray}
\bar \Lambda_{+,5/2} = 1.40 \mbox{ GeV} \, , \,
f_{+,5/2} = 0.20 \mbox{ GeV}^{9/2} \, ,
\end{eqnarray}
where the central value corresponds to $T=0.445$ GeV and $\omega_c = 3.0$ GeV.

The procedures are the same for different values of $\omega_c$. We give it
a large range $2.5$ GeV$<\omega_c<3.5$ GeV, but find that there are Borel windows
as long as $s_0 \geq 2.7$ GeV$^2$. The corresponding Borel windows and
the numerical results of $
\bar \Lambda_{+,5/2}$ and $f_{+,5/2}$ are listed
in Table.~\ref{tab:bw1}. We note that this table is shown in Sec.~\ref{sec:summary},
where we shall fix $\omega_c$ to evaluate $m_{2,+,5/2}$ and $m_{3,+,5/2}$.

\begin{figure}[hbtp]
\begin{center}
\begin{tabular}{c}
\scalebox{0.597}{\includegraphics{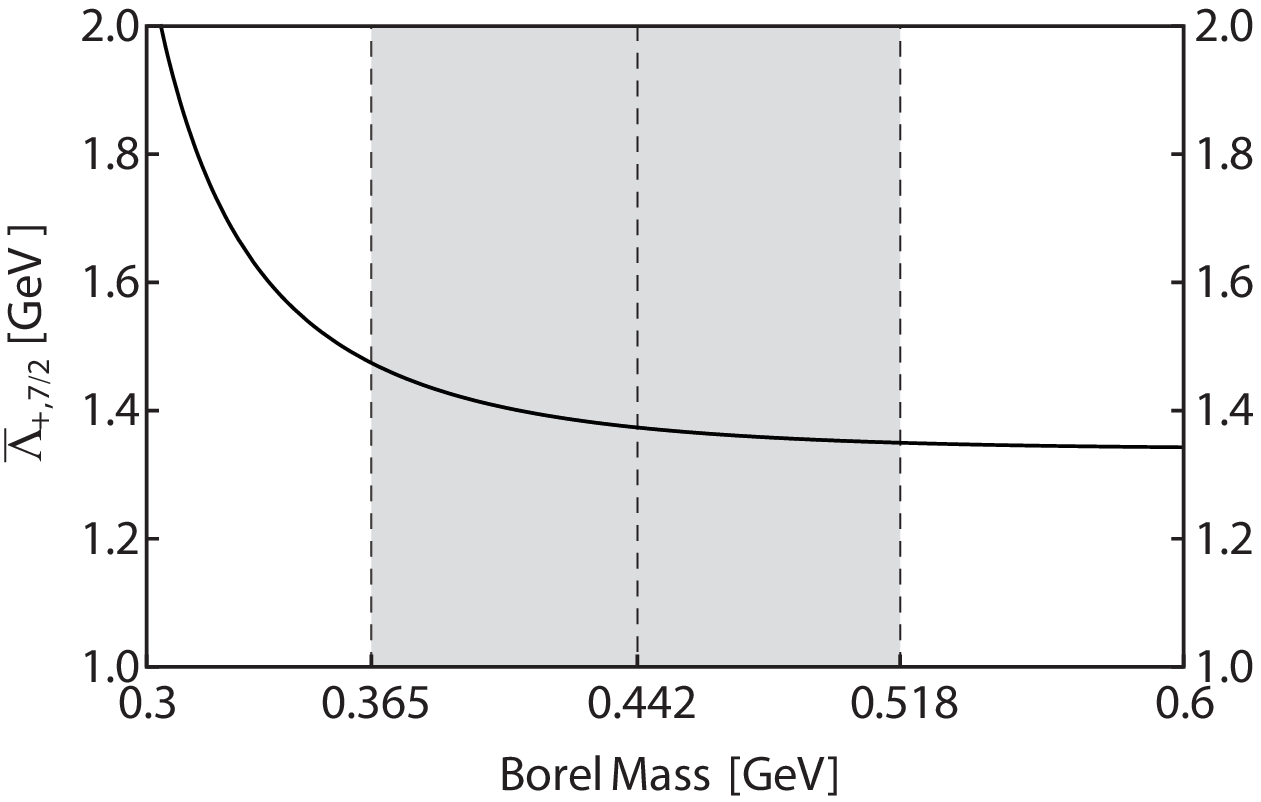}}\\
\scalebox{0.6}{\includegraphics{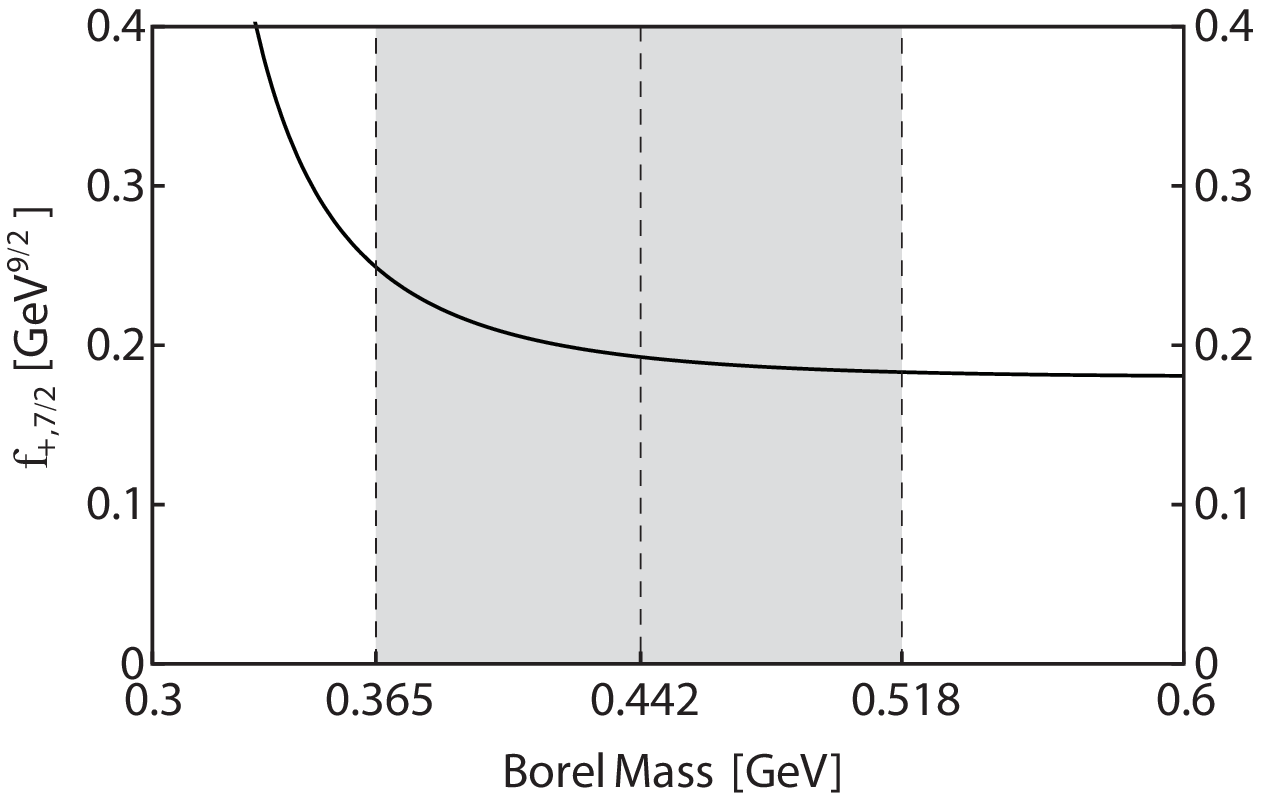}}
\end{tabular}
\caption{The
variations of $\bar \Lambda_{+,7/2}$ (top) and $f_{+,7/2}$ (bottom) with respect to the
Borel mass $T$.
In both figures we take $\omega_c$ = 3.0 GeV and the Borel window is $0.365$ GeV $< T < 0.518$ GeV.}
\label{fig:lambda2}
\end{center}
\end{figure}

Similarly, we use the sum rule (\ref{eq:ope2}) to perform QCD sum rule analyses.
The Borel windows and the numerical results of $\bar \Lambda_{+,7/2}$ and $f_{+,7/2}$ for various values
of $\omega_c$ are listed in Table~\ref{tab:bw2},
also shown in Sec.~\ref{sec:summary}. Here we show the variations of $\bar \Lambda_{+,7/2}$
and $f_{+,7/2}$ with respect to the
Borel mass $T$ in Fig.~\ref{fig:lambda2},
when we take $\omega_c = 3.0$ GeV and the Borel window is obtained to be $0.365$ GeV $< T < 0.518$ GeV. Again
these curves are more stable inside this window.
We obtain the following numerical results:
\begin{eqnarray}
\bar \Lambda_{+,7/2} = 1.37 \mbox{ GeV} \, , \,
f_{+,7/2} = 0.19 \mbox{ GeV}^{9/2} \, ,
\end{eqnarray}
where the central value corresponds to $T=0.442$ GeV and $\omega_c = 3.0$ GeV.

\section{The Sum Rules at the ${\mathcal O}(1/m_Q)$ Order}
\label{sec:nexttoleading}

In the previous section we have calculated $\bar \Lambda_{P,j_l} \equiv \lim_{m_Q \rightarrow \infty} (m_{j,P,j_l} - m_Q)$,
the value of which is the same for both $\bar \Lambda_{j_l-1/2,P,j_l}$ and $\bar \Lambda_{j_l+1/2,P,j_l}$.
To differentiate the masses within the same doublet, i.e., between $m_{j_l-1/2,P,j_l}$ and $m_{j_l+1/2,P,j_l}$, we need to work at
the $O(1/m_Q)$ order, which will be done in this section.
Again we follow the procedures used in Ref.~\cite{Zhou:2014ytp} (see Refs.~\cite{Zhou:2014ytp,Dai:1993kt,Dai:1996yw,Dai:1996qx,Dai:2003yg} for details),
and write the pole term on the hadron side,
Eq.~(\ref{eq:pole}), as:
\begin{eqnarray}
\Pi(\omega)_{pole} &=& \frac{(f+\delta
f)^{2}}{2(\overline{\Lambda}+\delta m)-\omega}
\nonumber\\ &=& \frac{f^{2}}{2\overline{\Lambda}-\omega}-\frac{2\delta mf^{2}}{(2\overline{\Lambda}-\omega)^{2}}+\frac{2f\delta f}{2\overline{\Lambda}-\omega} \, ,\label{eq:correction}
\end{eqnarray}
where we have omitted the subscripts ${j,P,j_l}$ for simplicity.
The corrections to the mass $m_{j,P,j_l}$ can be evaluated through
\begin{eqnarray}
\label{eq:next}
\delta m_{j,P,j_{l}} &=& -\frac{1}{4m_{Q}}\left(K_{P,j_{l}} + d_{j,j_l}C_{mag}\Sigma_{P,j_{l}} \right) \, ,
\end{eqnarray}
where $d_{j_{l}-1/2,j_{l}} = 2j_{l}+2$, $d_{j_{l}+1/2,j_{l}} = -2j_{l}$, and $C_{mag} (m_{Q}/\mu) = [ \alpha_s(m_Q) / \alpha_s(\mu)
]^{3/\beta_0}$ with $\beta_0 = 11 - 2 n_f /3$. The two corrections $K_{P,j_{l}}$ and $\Sigma_{P,j_{l}}$ come from the nonrelativistic kinetic
energy and the chromomagnetic interaction, respectively. We can calculate
them using the method of QCD sum rule in the
framework of HQET. We obtain the following two equations
for $K_{+,5/2}$ and $\Sigma_{+,5/2}$:
\begin{eqnarray}
&&f_{+,5/2}^2 K_{+,5/2} e^{-2 \bar \Lambda_{+,5/2} / T}
\label{eq:Kc1} \nonumber\\&&=  \int_{2
m_s}^{\omega_c} \left[ -{1 \over 9216 \pi^2} \omega^{10} + {161 {\langle g_s^2 GG \rangle} \over 30720 \pi^2}
w^6 \right] e^{-\omega/T} d\omega \, ,
\\&& f_{+,5/2}^2 \Sigma_{+,5/2} e^{-2 \bar \Lambda_{+,5/2} / T}
\label{eq:Sc1} = \int_{2
m_s}^{\omega_c} \left[ {{\langle g_s^2 GG \rangle} \over 25600 \pi^2}
w^6\right ] e^{-\omega/T} d\omega
\, ,\nonumber\\
\end{eqnarray}
and the following two equations for $K_{+,7/2}$ and $\Sigma_{+,7/2}$:
\begin{eqnarray}
&&f_{+,7/2}^2 K_{+,7/2} e^{-2 \bar \Lambda_{+,7/2} / T}
\label{eq:Kc2} \nonumber\\&&= \int_{2
m_s}^{\omega_c}\left [ -{3 \over 35840 \pi^2} \omega^{10} + {211 {\langle g_s^2 GG \rangle} \over 61440 \pi^2}
w^6 \right] e^{-\omega/T} d\omega \, ,
\\ &&f_{+,7/2}^2 \Sigma_{+,7/2} e^{-2 \bar \Lambda_{+,7/2} / T}
\label{eq:Sc2} = \int_{2
m_s}^{\omega_c}\left [ {{\langle g_s^2 GG \rangle} \over 26800 \pi^2}
w^6 \right] e^{-\omega/T} d\omega \, .\nonumber\\
\end{eqnarray}
Again, these sum rules for $(2^+,3^+)$ and $(3^+,4^+)$ are similar.
Then $K_{+,5/2}$, $\Sigma_{+,5/2}$, $K_{+,7/2}$, and $\Sigma_{+,7/2}$
can be simply obtained by dividing these equations with respect to the sum rules
(\ref{eq:ope1}) and (\ref{eq:ope2}). We evaluate their numerical results
in the Borel windows derived in the previous section, and list them for various values
of $\omega_c$ in Tables~\ref{tab:bw1} and \ref{tab:bw2}.

\begin{figure}[hbtp]
\begin{center}
\begin{tabular}{c}
\scalebox{0.6}{\includegraphics{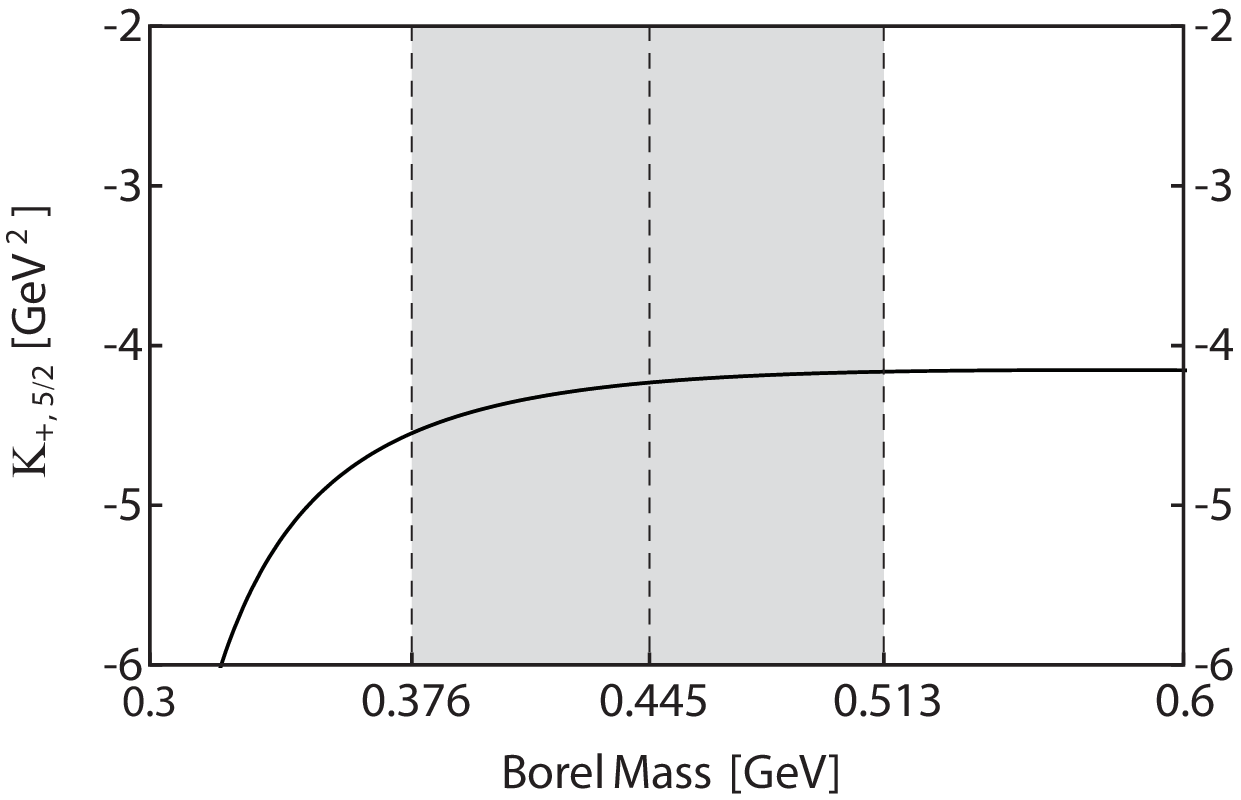}}\\
\scalebox{0.571}{\includegraphics{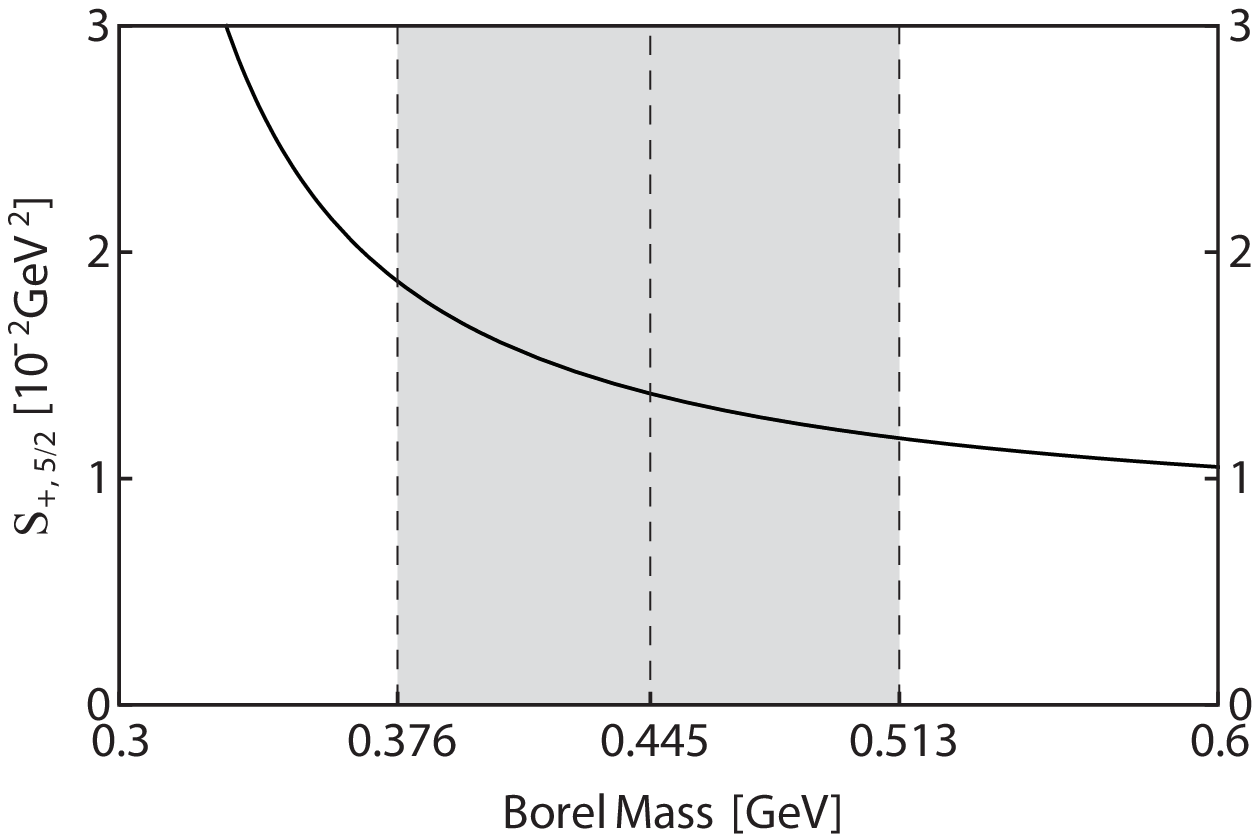}}
\end{tabular}
\caption{The variation
of $K_{+,5/2}$ (top) and $\Sigma_{+,5/2}$ (bottom) with respect to the Borel mass
$T$. In both figures we take $\omega_c$ = 3.0 GeV and
the Borel window is $0.376$ GeV $< T < 0.513$ GeV.} \label{fig:KS1}
\end{center}
\end{figure}

\begin{figure}[hbtp]
\begin{center}
\begin{tabular}{c}
\scalebox{0.6}{\includegraphics{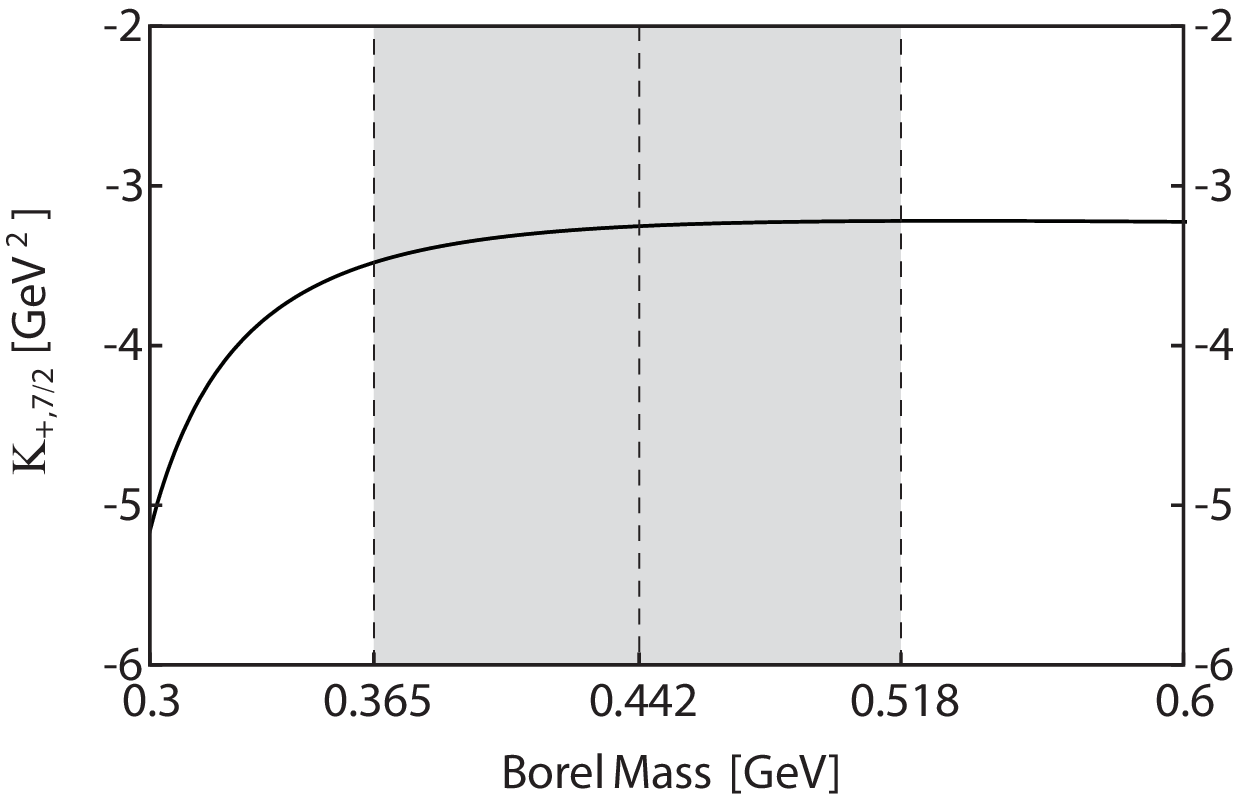}}\\
\scalebox{0.571}{\includegraphics{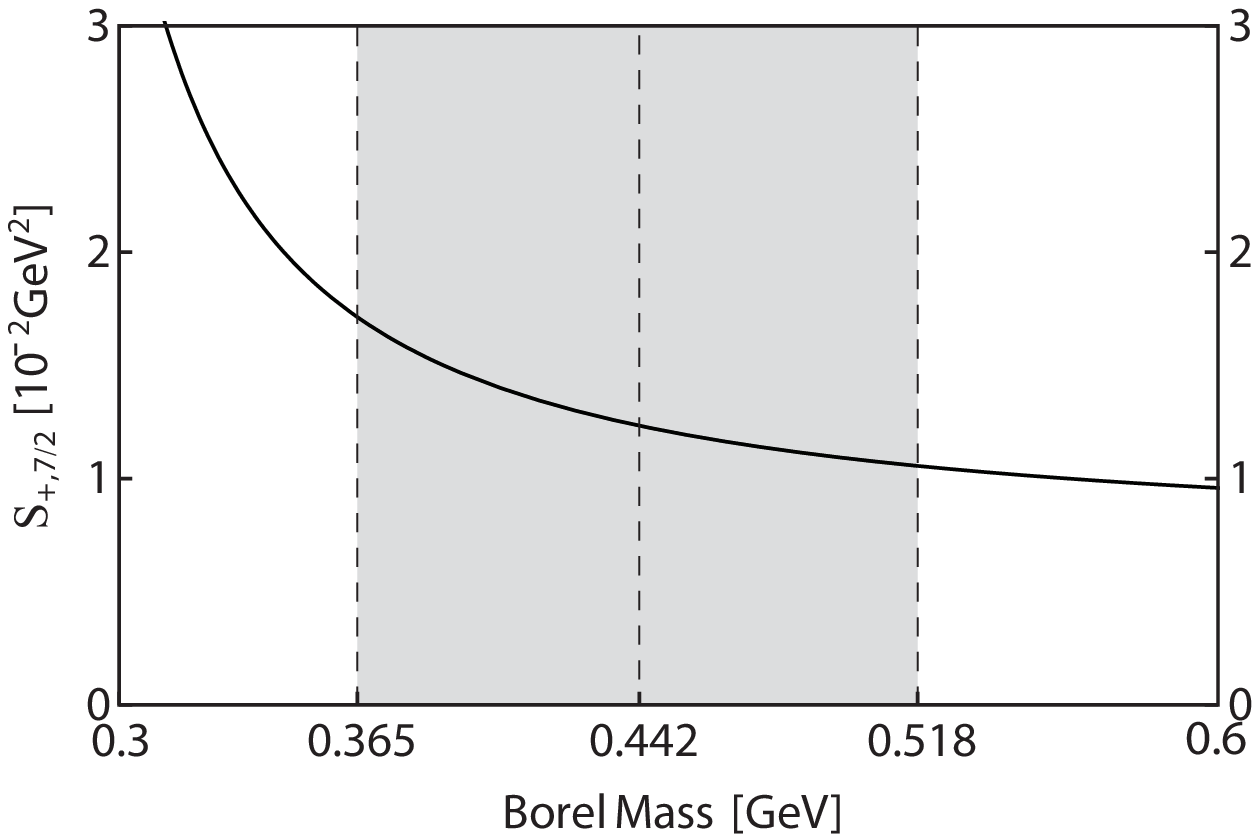}}
\end{tabular}
\caption{The variation
of $K_{+,7/2}$ (top) and $\Sigma_{+,7/2}$ (bottom) with respect to the Borel mass
$T$. In both figures we take $\omega_c$ = 3.0 GeV and
the Borel window is $0.365$ GeV $< T < 0.518$ GeV.} \label{fig:KS2}
\end{center}
\end{figure}

Here we take $\omega_c = 3.0$ GeV as an example, and show their variations with respect to the
Borel mass $T$ in Figs.~\ref{fig:KS1} and \ref{fig:KS2}. We use the Borel windows $0.376$ GeV $< T < 0.513$ GeV
for $K_{+,5/2}$ and $\Sigma_{+,5/2}$, and obtain the following numerical results:
\begin{eqnarray}
K_{+,5/2} = -4.23 \mbox{ GeV}^2 \, , \,
\Sigma_{+,5/2} = 0.014 \mbox{ GeV}^{2} \, ,
\end{eqnarray}
where the central value corresponds to $T=0.445$ GeV and $\omega_c = 3.0$ GeV.
We use the same Borel window $0.365$ GeV $< T < 0.518$ GeV for $K_{+,7/2}$ and $\Sigma_{+,7/2}$,
and obtain the following numerical results:
\begin{eqnarray}
K_{+,7/2} = -3.25 \mbox{ GeV}^2 \, , \,
\Sigma_{+,7/2} = 0.012 \mbox{ GeV}^{2} \, ,
\end{eqnarray}
where the central value corresponds to $T=0.442$ GeV and $\omega_c =
3.0$ GeV.

\section{Numerical Results and Discussions}
\label{sec:summary}

The mass of the $F$-wave $\bar c s$ heavy mesons can be obtained using
Eqs.~(\ref{eq:lambda}) and (\ref{eq:next}). We use $(m_{D^*_{s2}},m_{D_{s3}})$
to denote the mass of the heavy mesons belonging to the
$(2^+,3^+)$ spin doublet, and they satisfy:
\begin{eqnarray}
{1\over12} ( 5 m_{D^*_{s2}} + 7 m_{D_{s3}} ) &=& m_c + \bar \Lambda_{+,5/2} - {1 \over 4 m_c} K_{+,5/2} \, ,\nonumber\\
\label{eq:convergence1}
\\
m_{D_{s3}} - m_{D^*_{s2}} &=& {3 \over m_c} \Sigma_{+,5/2} \, .
\end{eqnarray}
We use $(m_{D^\prime_{s3}},m_{D^*_{s4}})$ to denote the mass of the heavy mesons belonging to the
$(3^+,4^+)$ spin doublet, and they satisfy:
\begin{eqnarray}
{1\over16} ( 7 m_{D^\prime_{s3}} + 9 m_{D^*_{s4}} ) &=& m_c + \bar \Lambda_{+,7/2} - {1 \over 4 m_c} K_{+,7/2} \, ,\nonumber\\
\label{eq:convergence2}
\\
m_{D^*_{s4}} - m_{D^\prime_{s3}} &=& {4 \over m_c} \Sigma_{+,7/2} \, .
\end{eqnarray}
In this paper we use the charm quark mass $m_c = 1.275 \pm 0.025$ GeV, which is evaluated in the $\overline{\rm MS}$ scheme~\cite{Agashe:2014kda}.
Using the above equations, we calculate $(m_{D^*_{s2}},m_{D_{s3}})$, $(m_{D^\prime_{s3}},m_{D^*_{s4}})$, and their differences for various threshold values $\omega_c$. The results are listed
in Tables~\ref{tab:bw1} and \ref{tab:bw2}. For completeness, we also list Borel windows, $\bar \Lambda_{P,j_l}$, $f_{P,j_l}$, $K_{P,j_l}$, and $\Sigma_{P,j_l}$ for various $\omega_c$.

\begin{figure}[hbtp]
\begin{center}
\begin{tabular}{c}
\scalebox{0.597}{\includegraphics{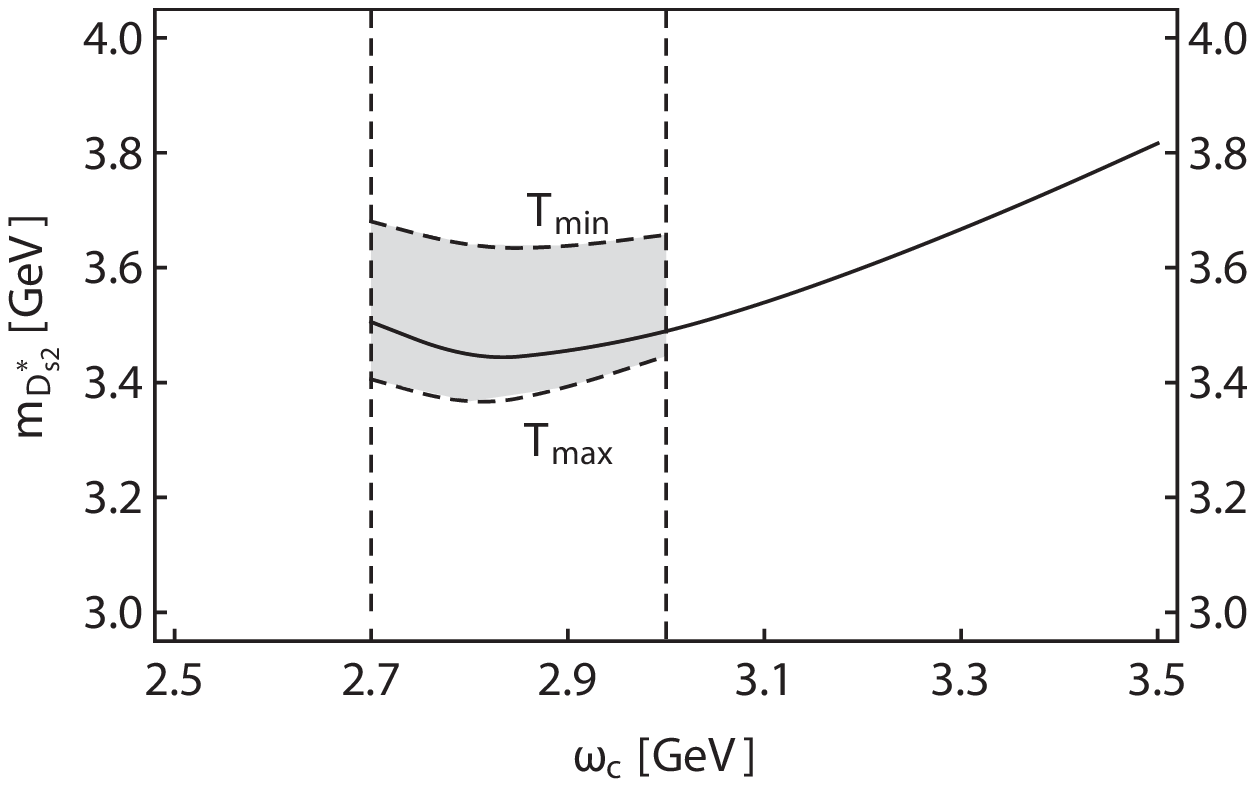}}\\
\scalebox{0.615}{\includegraphics{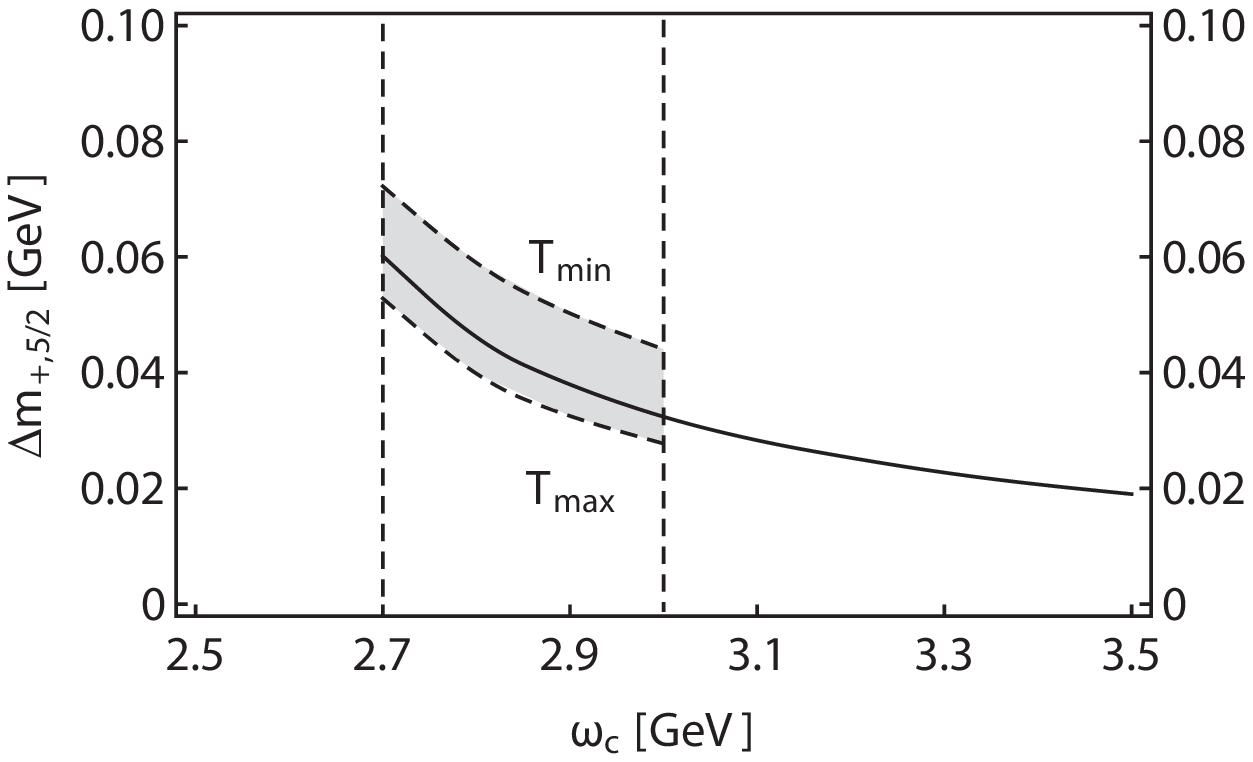}}
\end{tabular}
\caption{The variations of $m_{D^*_{s2}}$ (top) and $\Delta m_{+,5/2}$ (bottom) with respect to the
threshold values $\omega_c$. The upper and lower bands are obtained by using $T_{min}$ and $T_{max}$, respectively.}
\label{fig:mass52}
\end{center}
\end{figure}

\begin{figure}[hbtp]
\begin{center}
\begin{tabular}{c}
\scalebox{0.597}{\includegraphics{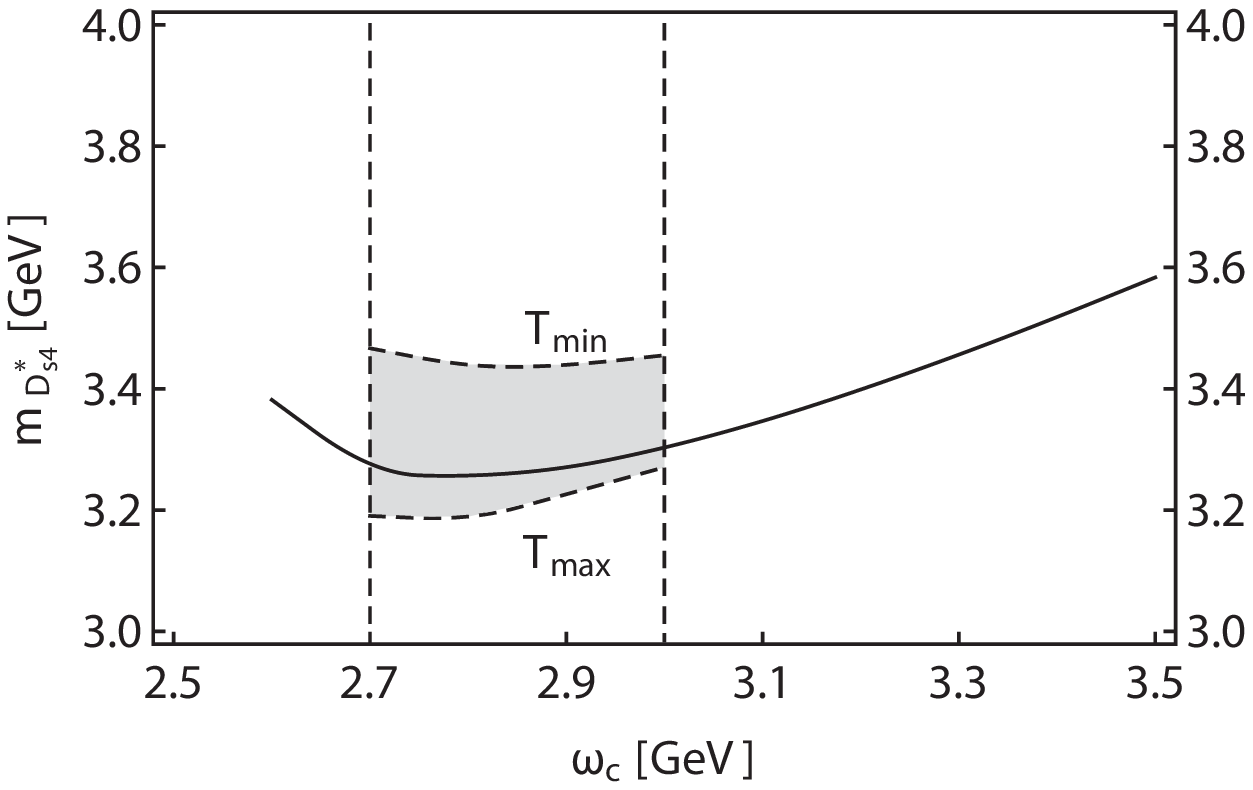}}\\
\scalebox{0.613}{\includegraphics{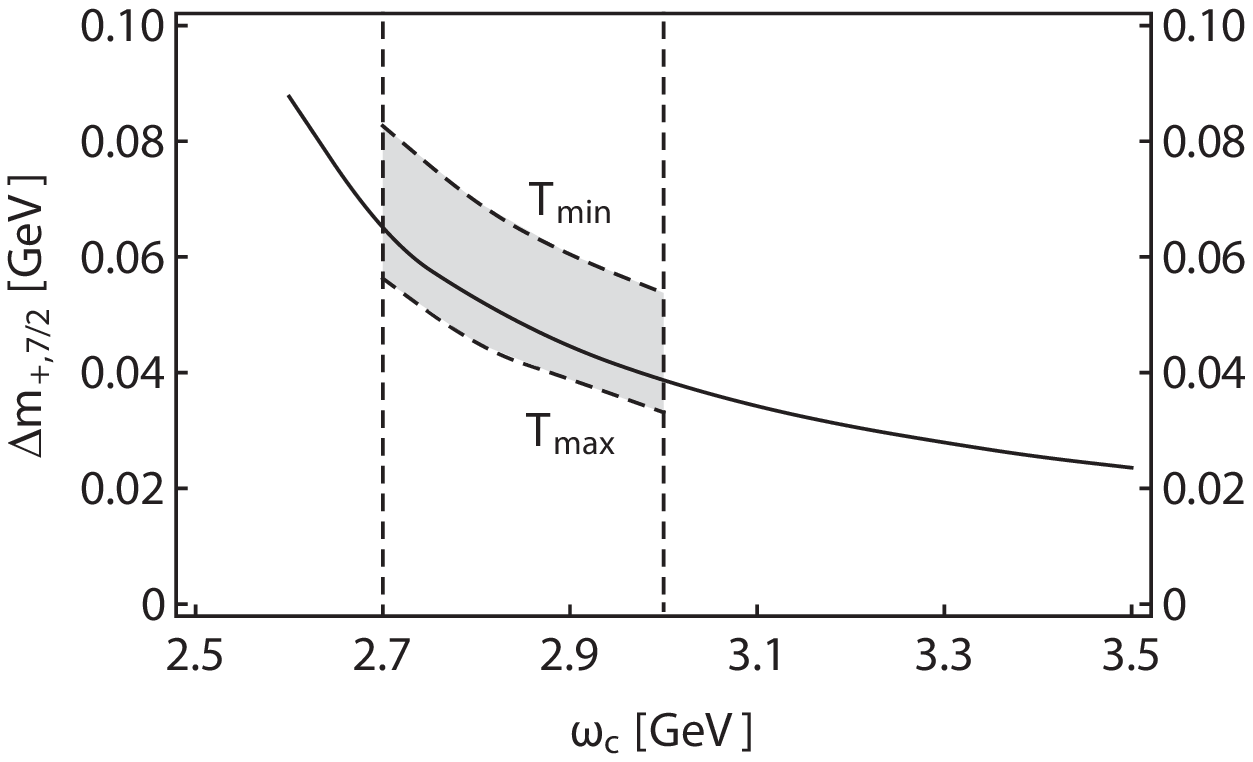}}
\end{tabular}
\caption{The variations of $m_{D^*_{s4}}$ (top) and $\Delta m_{+,7/2}$ (bottom) with respect to the
threshold values $\omega_c$. The upper and lower bands are obtained by using $T_{min}$ and $T_{max}$, respectively.}
\label{fig:mass72}
\end{center}
\end{figure}

Now we can fix the the threshold value $\omega_c$. Our criterion is to require that the $\omega_c$ dependence of
the mass prediction be the weakest. We show variations of $m_{D^*_{s2}}$ and $m_{D^*_{s4}}$
with respect to the threshold value $\omega_c$ in the top panels of Figs.~\ref{fig:mass52} and \ref{fig:mass72},
and quickly notice that this dependence is the weakest around $\omega_c \sim 2.8$ GeV for both cases.
Accordingly, we choose the region $2.7$ GeV$<\omega_c<3.0$ GeV as our working region. We obtain
the following numerical results for the $(2^+,3^+)$ spin doublet:
\begin{eqnarray}
\nonumber m_{D^*_{s2}} &=& 3.45 \pm 0.25 \mbox{ GeV} \, , \,
\\ m_{D_{s3}} &=& 3.50 \pm 0.26 \mbox{ GeV} \, , \,
\\ \nonumber \Delta m_{+,5/2} &=& 0.046 \pm 0.030 \mbox{ GeV} \, ,
\end{eqnarray}
where the central value corresponds to $T=0.418$ GeV and $\omega_c = 2.8$ GeV. Here the
uncertainties are due to the
Borel mass $T$, the threshold value $\omega_c$, and the uncertainty of
the gluon condensate $\langle {\alpha_s\over\pi} GG\rangle = 0.005 \pm 0.001 \mbox{ GeV}^4$. We obtain
the following numerical results for the $(3^+,4^+)$ spin doublet
\begin{eqnarray}
\nonumber m_{D^\prime_{s3}} &=& 3.20 \pm 0.22 \mbox{ GeV} \, , \,
\\ m_{D^*_{s4}} &=& 3.26 \pm 0.23 \mbox{ GeV} \, , \,
\\ \nonumber \Delta m_{+,7/2} &=& 0.053 \pm 0.044 \mbox{ GeV} \, ,
\end{eqnarray}
where the central value corresponds to $T=0.417$ GeV and $\omega_c = 2.8$ GeV.
However, we note that the mass differences within the same doublets, $\Delta m_{+,5/2}= m_{D_{s3}} - m_{D^*_{s2}}$ and $\Delta m_{+,7/2}= m_{D^*_{s4}} - m_{D^\prime_{s3}}$,
do depend on the threshold value $\omega_c$, as shown in the bottom panels of Figs.~\ref{fig:mass52} and \ref{fig:mass72}.

\renewcommand{\arraystretch}{1.5}
\begin{table*}[hbtp]
\begin{center}
\caption{The mass of the heavy mesons belonging to the $(2^+,3^+)$ spin doublet $m_{D^*_{s2}}$ and $m_{D_{s3}}$, and their differences $\Delta m_{+,5/2}= m_{D_{s3}} - m_{D^*_{s2}}$, for various threshold values $\omega_c$. We also
list Borel windows, $\bar \Lambda_{+,5/2}$, $f_{+,5/2}$, $K_{+,5/2}$, and $\Sigma_{+,5/2}$ for completeness. }
\begin{tabular}{cccccccccc}
\toprule[1pt]\toprule[1pt]
\mbox{$\omega_c [\rm GeV]$} & \mbox{Borel window[GeV]} & \mbox{$\bar \Lambda [\rm GeV]$} & \mbox{$f[\rm GeV^{9/2}]$} & \mbox{$K[\rm GeV^2]$} & \mbox{$\Sigma[\rm GeV^2]$} & \mbox{$m_{D^*_{s2}}[\rm GeV]$} & \mbox{$m_{D_{s3}}[\rm GeV]$} & \mbox{$\Delta m [\rm GeV]$}
\\ \midrule[1pt]
2.7 & [0.376,0.426] & 1.4835 & 0.1748 & -3.9886 & 0.02557 & 3.5055 & 3.5656 & 0.0601
\\
2.8 & [0.376,0.459] & 1.4210 & 0.1701 & -3.9779 & 0.01961 & 3.4491 & 3.4953 & 0.0462
\\ 
2.9 & [0.376,0.488] & 1.4028 & 0.1812 & -4.0794 & 0.01613 & 3.4556 & 3.4935 & 0.0379
\\ 
3.0 & [0.376,0.513] & 1.4036 & 0.1992 & -4.2317 & 0.01375 & 3.4895 & 3.5218 & 0.0323
\\
3.1 & [0.376,0.537] & 1.4151 & 0.2221 & -4.4156 & 0.01203 & 3.5394 & 3.5677 & 0.0283
\\ 
3.2 & [0.376,0.560] & 1.4333 & 0.2491 & -4.6213 & 0.01071 & 3.5997 & 3.6250 & 0.0253
\\ 
3.3 & [0.376,0.582] & 1.4555 & 0.2800 & -4.8432 & 0.009658 & 3.6669 & 3.6897 & 0.0228
\\
3.4 & [0.376,0.603] & 1.4807 & 0.3146 & -5.0789 & 0.008789 & 3.7395 & 3.7602 & 0.0207
\\ 
3.5 & [0.376,0.624] & 1.5081 & 0.3532 & -5.3259 & 0.008076 & 3.8163 & 3.8353 & 0.0190
\\ \bottomrule[1pt]\bottomrule[1pt]
\end{tabular}
\label{tab:bw1}
\end{center}
\end{table*}

\begin{table*}[hbtp]
\begin{center}
\caption{The mass of the heavy mesons belonging to the $(3^+,4^+)$ spin doublet $m_{D^\prime_{s3}}$ and $m_{D^*_{s4}}$, and their differences $\Delta m_{+,7/2}= m_{D^*_{s4}} - m_{D^\prime_{s3}}$, for various threshold values $\omega_c$. We also
list Borel windows, $\bar \Lambda_{+,7/2}$, $f_{+,7/2}$, $K_{+,7/2}$, and $\Sigma_{+,7/2}$ for completeness.}
\begin{tabular}{ccccccccc}
\toprule[1pt]\toprule[1pt]
\mbox{$\omega_c [\rm GeV]$} & \mbox{Borel window[GeV]} & \mbox{$\bar \Lambda [\rm GeV]$} & \mbox{$f[\rm GeV^{9/2}]$} & \mbox{$K[\rm GeV^2]$} & \mbox{$\Sigma[\rm GeV^2]$} & \mbox{$m_{D^\prime_{s3}}[\rm GeV]$} & \mbox{$m_{D^*_{s4}}[\rm GeV]$} & \mbox{$\Delta m [\rm GeV]$}
\\ \midrule[1pt]
2.6 & [0.365,0.404] & 1.4690 & 0.1594 & -3.0566 & 0.02796 & 3.2940 & 3.3817 & 0.0877
\\ 
2.7 & [0.365,0.440] & 1.3884 & 0.1486 & -2.9815 & 0.02076 & 3.2114 & 3.2765 & 0.0651
\\ 
2.8 & [0.365,0.469] & 1.3649 & 0.1571 & -3.0294 & 0.01682 & 3.2043 & 3.2570 & 0.0527
\\ 
2.9 & [0.365,0.495] & 1.3632 & 0.1725 & -3.1262 & 0.01420 & 3.2262 & 3.2707 & 0.0445
\\
3.0 & [0.365,0.518] & 1.3735 & 0.1926 & -3.2523 & 0.01234 & 3.2645 & 3.3032 & 0.0387
\\
3.1 & [0.365,0.542] & 1.3907 & 0.2165 & -3.3976 & 0.01090 & 3.3126 & 3.3469 & 0.0343
\\ 
3.2 & [0.365,0.564] & 1.4127 & 0.2442 & -3.5579 & 0.009792 & 3.3680 & 3.3988 & 0.0308
\\ 
3.3 & [0.365,0.585] & 1.4379 & 0.2754 & -3.7294 & 0.008901 & 3.4284 & 3.4563 & 0.0279
\\ 
3.4 & [0.365,0.606] & 1.4651 & 0.3102 & -3.9120 & 0.008126 & 3.4928 & 3.5183 & 0.0255
\\
3.5 & [0.365,0.627] & 1.4941 & 0.3488 & -4.1013 & 0.007510 & 3.5600 & 3.5836 & 0.0236
\\ \bottomrule[1pt]\bottomrule[1pt]
\end{tabular}
\label{tab:bw2}
\end{center}
\end{table*}

\begin{table*}[hbtp]
\begin{center}
\caption{Masses of $F$-wave charmed-strange ($\bar c s$) mesons (in GeV).}
\label{table:mass}
\begin{tabular}{cc|ccccc}
\toprule[1pt]\toprule[1pt]
\mbox{State} & \mbox{This work} & \mbox{State} & \mbox{Ref.~\cite{Song:2015nia}} & \mbox{Ref.~\cite{Ebert:2009ua}} & \mbox{Ref.~\cite{Di Pierro:2001uu}}
\\ \midrule[1pt]
$2^+$ in $(2^+,3^+)$ & $3.45 \pm 0.25$ & $1^3F_2$ &  3.159  & 3.230 & 3.224
\\
$3^+$ in $(2^+,3^+)$ & $3.50 \pm 0.26$ & $1^1F_3$ &  3.151  & 3.266 & 3.247
\\
$3^+$ in $(3^+,4^+)$ & $3.20 \pm 0.22$ & $1^3F_3$ &  3.157  & 3.254 & 3.203
\\
$4^+$ in $(3^+,4^+)$ & $3.26 \pm 0.23$ & $1^3F_4$ &  3.143  & 3.300 & 3.220
\\ \bottomrule[1pt]\bottomrule[1pt]
\end{tabular}
\label{tab:bw3}
\end{center}
\end{table*}

\begin{figure}[hbtp]
\begin{center}
\scalebox{0.597}{\includegraphics{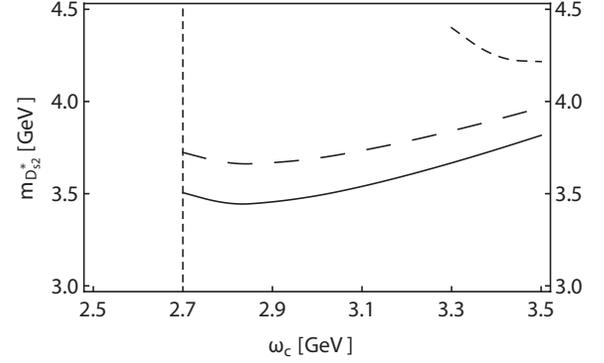}}
\caption{The variations of $m_{D^*_{s2}}$ with respect to the
threshold values $\omega_c$.
The solid curve is our previous result which has been plotted in the top panel of Fig.~\ref{fig:mass52};
the short-dashed curve is obtained by using the gluon condensate $\langle {\alpha_s\over\pi} GG\rangle =0.012$ GeV$^4$~\cite{Shifman:1978bx};
the long-dashed curve is obtained by using the pole mass of the charm quark $m_c = 1.67$~GeV~\cite{Agashe:2014kda}.}
\label{fig:massDs2}
\end{center}
\end{figure}

The above analyses suggest that there is a heavy meson spin doublet $(2^+,3^+)$ whose masses
are around 3.45 GeV and 3.50 GeV, and a spin doublet $(3^+,4^+)$ whose masses are around 3.20 GeV and 3.26 GeV. The latter is consistent with recent
theoretical studies~\cite{Song:2015nia,Ebert:2009ua,Di Pierro:2001uu}, while the former is larger but still within uncertainties, as shown in Table~\ref{table:mass}.
We note that the two sum rules for $(2^+,3^+)$ and $(3^+,4^+)$ are similar, see Eqs.~(\ref{eq:ope1}) and (\ref{eq:ope2}), Eqs.~(\ref{eq:Kc1}) and (\ref{eq:Kc2}),
and Eqs.~(\ref{eq:Sc1}) and (\ref{eq:Sc2}), so the mass difference between $(2^+,3^+)$ and $(3^+,4^+)$ may be (partly) due to the theoretical uncertainty of
the numerical analysis. Moreover, the expansion on the charm quark mass for the $(2^+,3^+)$ spin doublet is
\begin{eqnarray}
{\rm Eq.}~(\ref{eq:convergence1}) \sim m_c + 1.4 {\rm ~GeV} + 1.0 {\rm ~GeV} \, ,
\end{eqnarray}
while the expansion for the $(3^+,4^+)$ spin doublet has better convergence
\begin{eqnarray}
{\rm Eq.}~(\ref{eq:convergence2}) \sim m_c + 1.4 {\rm ~GeV} + 0.8 {\rm ~GeV} \, ,
\end{eqnarray}
This suggests that our results for the latter doublet are more reliable.

To make our analyses complete, we try to change the values of the parameters used in the previous analyses and redo the calculations:
\begin{enumerate}

\item As shown in sum rules (\ref{eq:ope1}) and (\ref{eq:ope2}), the gluon condensate is important. Besides the value
listed in Eqs.~(\ref{condensates}), $\langle {\alpha_s\over\pi} GG\rangle = 0.005 \pm 0.004$ GeV$^4$~\cite{Ioffe:2005ym,Narison:2002pw}, the value
$\langle {\alpha_s\over\pi} GG\rangle =0.012\pm 0.004$ GeV$^4$ is also widely used in QCD sum rule studies~\cite{Shifman:1978bx} (see Ref.~\cite{Ioffe:2005ym,Narison:2002pw}
for detailed discussions). We use this value and redo the numerical analyses. The mass of the $2^+$ heavy meson, $m_{D^*_{s2}}$, is shown in Fig.~\ref{fig:massDs2} with
respective to the threshold value $\omega_c$, using short-dashed curves. The obtained result is even larger than 4.0 GeV, which is not very reliable/reasonable.

\item We change the charm quark mass from the $\overline{\rm MS}$ value $m_c = 1.275 \pm 0.025$ GeV to its pole mass $m_c = 1.67 \pm 0.07$~\cite{Agashe:2014kda}, and redo the numerical analyses.
The result is shown in Fig.~\ref{fig:massDs2} with respective to $\omega_c$ using long-dashed curve. The obtained result is about 200 MeV larger
than our previous result, suggesting that our results for the masses of the heavy mesons can have significant
theoretical uncertainties (see also discussions in Ref.~\cite{Zhou:2014ytp}).

\end{enumerate}

We can similarly replace the charm quark by bottom quark and study the $\bar bs$ system (the factor $C_{mag}$ in Eq.~(\ref{eq:next}) is taken to be 0.8~\cite{Dai:1996qx,Dai:2003yg}).
Again, these masses depend much on the bottom quark mass $m_b$, whose value has large uncertainties.
When we use the ${\rm 1S}$ mass value $m_b = 4.66$ GeV~\cite{Agashe:2014kda}, we can obtain the mass of the $F$-wave $\bar bs$ heavy-light mesons
to be around 6.3 GeV, consistent with the results obtained in Ref.~\cite{Ebert:2009ua,Sun:2014wea}.
Their mass differences are $\Delta m^{[\bar bs]}_{+,5/2} \sim 0.010$ GeV and $\Delta m^{[\bar bs]}_{+,7/2} \sim 0.014$ GeV.
However, if we replace the strange quark by up and down quarks, the sum rules (\ref{eq:ope1}) and (\ref{eq:ope2}) would become too
simple to investigate the non-strange $D$-wave heavy mesons.

In summary, in this work we adopt the QSR approach to study the mass spectrum of $F$-wave heavy-light mesons in the framework of HQET.
We obtain two similar sum rules for $(2^+,3^+)$ and $(3^+,4^+)$, see Eqs.~(\ref{eq:ope1}) and (\ref{eq:ope2}), Eqs.~(\ref{eq:Kc1}) and (\ref{eq:Kc2}),
and Eqs.~(\ref{eq:Sc1}) and (\ref{eq:Sc2}).
Our results suggest that there is a $\bar c s$ heavy meson spin doublet $(2^+,3^+)$ whose masses
are $m^{[\bar c s]}_{(2^+,3^+)}$ = ($ 3.45 \pm 0.25$, $3.50 \pm 0.26$) GeV, with mass difference $\Delta m^{[\bar c s]}_{+,5/2} = 0.046 \pm 0.030$ GeV, and a spin doublet $(3^+,4^+)$ whose masses are $m^{[\bar c s]}_{(3^+,4^+)}$ = ($3.20 \pm 0.22$, $3.26 \pm 0.23$) GeV,
with mass difference $\Delta m^{[\bar c s]}_{+,7/2} = 0.053 \pm 0.044$ GeV.
We note that this is a pioneering work and these results are provisional.

Finally, we would like to note that this is a pioneering study applying HQET-based QSR to study $F$-wave heavy-light mesons (see also discussions in Sec.~\ref{sec:intro}). They have large orbital excitations $L=3$, which can be explicitly written as three covariant derivatives, and are not easy to deal with. However, because the LHCb experiments have just observed $D$-wave heavy mesons~\cite{Aaij:2014xza,Aaij:2014baa}, the theoretical analyses on $F$-wave heavy mesons, including our study in current paper, become helpful to the further experimental exploration of them. Moreover, they are also good challenges (tests) for theoretical studies.
In the following experiments such as LHCb and BelleII, searching for higher excitations of heavy-light mesons will be an important task. We expect more experimental and theoretical progresses on higher excitations of heavy-light mesons, which will make our knowledge of heavy-light meson family become more and more abundant. This will improve our understanding to the non-perturbative behavior of QCD, and inspire ideas for the improvement of QCD sum rule itself.

\vfil

\section*{Acknowledgments}

We would like to thank the anonymous referee for his/her valuable suggestion.
This project is supported by the National Natural Science Foundation
of China under Grants No. 11205011, No. 11475015, No. 11375024, No. 11222547, No.
11175073, No. 11035006, and NO. 11261130311, the Ministry of
Education of China (SRFDP under Grant No. 20120211110002 and the
Fundamental Research Funds for the Central Universities), and the
Fok Ying-Tong Education Foundation (No. 131006).
\vfil

\end{document}